\documentstyle[prb,aps,epsfig,float]{revtex}

\begin{document}
\title{Dependence of the physical properties of
$Nd_{0.5}Ca_{0.5}MnO_{3+\delta}$ on the oxydation state of $Mn$.}

\author{Carlos Frontera and Jos\'e Luis Garc\'{\i}a-Mu\~noz}
\address{Institut de Ci\`encia de Materials de Barcelona, CSIC, Campus
de la UAB, E-08193 Bellaterra, Spain.}
\author{Anna Llobet}
\address{Laboratoire Louis N\'eel, 25 avenue des Martyrs, Polygone
CNRS BP 166, 38042 Grenoble CEDEX 09, France.} 
\author{Clemens Ritter}
\address{Institute Laue Lagevin, 6, rue Jules Horowitz, BP
156, 38042 Grenoble-Cedex 9, France.}
\author{Jos\'e Antonio Alonso}
\address{Instituto de Ciencia de Materiales de Madrid, CSIC,
Cantoblanco, 28049 Madrid, Spain.}%
\author{Juan Rodr\'{\i}guez-Carvajal} 
\address{Laboratoire L\'eon Brillouin,
CEA-CNRS, Centre d'Etude de Saclay, 91191 Gif sur Yvette Cedex,
France.} 
\date{\today}
\wideabs{ 
\maketitle

\begin{abstract}
We present a comparative study of the magnetic, transport and
structural properties of $Nd_{0.5}Ca_{0.5}MnO_{3+\delta}$
\protect{[}$\delta=$ 0.02(1) and 0.04(1)\protect{]}. We have found
significant differences between the low temperature magnetic orders
and the magnetization curves below $T_{CO}\approx 250 K$ of the two
samples. In particular one component of the magnetic moment presents a
ferromagnetic coupling between the $(0\;0\;1)$ planes ($P\,bnm$
setting) deviating the angle between neighboring $Mn$ ions from
$180^o$ (perfect CE order) to $150^o$
\protect{[}$\delta=0.02(1)$\protect{]} and $130^o$ \protect{[}$\delta=
0.04(1)$\protect{]}. These results imply a remarkable $\delta$
dependence which is discussed in the light of a non-random spatial
distribution of defects in the perfect charge order scheme.

\end{abstract}
}
\narrowtext


A considerable research effort concerning the charge order (CO)
phenomenon in doped manganites ($Ln_{1-x}A_xMnO_3$ with $Ln=$ rare
earth and $A=$ alkaline earth) has been made during the last years.
Ingredients such as Coulomb repulsion, the effect of Jahn-Teller
distortion on the $e_g$ energy levels and lattice distortions play an
important role on the stability of the CO state. This state is usually
accompanied by a real space ordering of the orbitals occupied by the
$e_g$ electrons. When this orbital order (OO) occurs, the
$Mn$-$O$-$Mn$ superexchange interactions are FM if a half-filled
$d_{3r_i^2-r^2}$ ($r_i=x,y$ or $z$) and an empty $e_{g}$
orbitals are involved and antiferromagnetic (AFM) if the involved ones
are two $t_{2g}$ orbitals.\cite{Wollan55,Radaelli97} This makes the
magnetic behavior to be strongly dependent on (i) the mean valence
state of the $Mn$ ions (or, equivalently, on the density of $e_g$
electrons) and (ii) the real space OO.  When the formal ratio
$\frac{Mn^{+4}}{Mn^{+3}}$ ions is 1, and for a rich variety of rare
earths the low temperature magnetic structure found is the
CE\cite{Wollan55}. This structure is usually explained as resulting
from a well established CO and OO.\cite{Radaelli97,Wollan55} Some
previous works have been devoted to study how the off-stoichiometry of
the $\frac{Mn^{+4}}{Mn^{+3}}=1$ ratio  affects the
CO.\cite{Roy98,Tomioka96,Barnabe98,Chen99,Richard99} One interesting
effect is the variation in the CO modulation wave vector
[$\makebox{\boldmath $Q$}=\frac {2\pi}a (0\;1/2-\alpha\;0)$ in the
$P\,bnm$ setting] that different annealing conditions produce, for example, in
$Pr_{0.5} Ca_{0.5} MnO_3$ and in $Sm_{0.5} Ca_{0.5}
MnO_3$.\cite{Barnabe98} Incommensurability of the CO in
$La_{0.5}Ca_{0.5}MnO_3$ have also been reported in the temperature
range of the structural transformation,  but it disappears at lower
temperature.\cite{Chen96} Barnab\'e {\em et
al.}~(Ref.~\onlinecite{Barnabe98}) argue that in an under-doped sample
(presenting CO) the extra $e_g$ electrons are randomly placed in the
$Mn$ positions leaving the CO commensurated ($\alpha= 0$), but the
extra $Mn^{+4}$ ions of an over-doped sample introduce
incommensurability in the CO structure ($\alpha > 0$) even when the
excess of $Mn^{+4}$ is very small. Using electron diffraction data,
Barnab\'e {\em et al.} conclude that the extra $Mn^{+4}$ ions form
ordered $(0\;1\;0)$ planes. Such an incommensurability of an
over-doped sample has been also reported by Chen {\em et al.} in
Ref.~\onlinecite{Chen99}.

In this brief report we present the effects of changing the oxydation
state  of $Mn$ in $Nd_{0.5}Ca_{0.5}MnO_{3+\delta}$. The notation we
use here is convenient from the point of view of the chemical
formulation, but the actual structural defects are not interstitial
oxygens, they should correspond to cation vacancies. This study is
based on the comparison of the structural, magnetic and transport
properties. The main changes correspond to the magnetic structure. Its
evolution with $\delta$ indicates that the extra holes due to $\delta$
are not randomly distributed.


Two different polycrystalline samples have been sintered by standard
solid state reaction by mixing high purity powders of $CaCO_{3}$,
$Mn_{2}O_3$ and $Nd_2O_3$ in appropiate ratios for the nominal
composition $Nd_{1/2}Ca_{1/2}MnO_{3}$. After some intermediate
treatments, the two samples were pressed into pellets, fired and
ground again for several times. For the first sample (air sample) the
firings were done at $1450^oC$ in air followed by a rapid quench to RT
($-500^oC/hour$). For the second sample ($O_2$ sample) the firings
were done at $1400^oC$ in an atmosphere of flowing $O_2$ and the final
one was followed by a slow cooling to RT ($-50^oC/hour$). X-ray powder
diffraction confirmed that both samples are well crystallized in a
single phase. Agreeing with the synthesis conditions,
thermogravimetric analysis (TGA), used to determine the oxygen content
of the samples ($Nd_{0.5}Ca_{0.5}MnO_{3+\delta}$), evidenced a larger
content of oxygen in the $O_2$ ($\delta = 0.04(1)$, $Mn^{+4}=
58(2)\%$) than in air sample ($\delta= 0.02(1)$, $Mn^{+4}=
54(2)\%$). Neutron diffraction (ND) patterns of the air sample were
collected at the Institute Laue Langevin (Grenoble) using D2B
($\lambda= 1.594\rm\AA$) and D1B ($\lambda=2.52\rm\AA$)
diffractometers. ND patterns of the $O_2$ sample were collected at the
Laboratoire Leon Brillouin (Paris) using 3T2 ($\lambda= 1.227\rm\AA$)
and G4.2 ($\lambda= 2.426\rm\AA$) diffractometers. For both samples,
ND patterns were collected for several temperatures in the range
$1.5K$ to room temperature (RT).  They were analyzed by the Rietveld
method using the program FULLPROF.\cite{JRC93} Resistivity was
measured by the four-probe method using a commercial PPMS (Quantum
Design).  Magnetization measurements have been carried out using a
commercial SQUID (Quantum Design).


\begin{figure}
\centerline{\epsfig{file=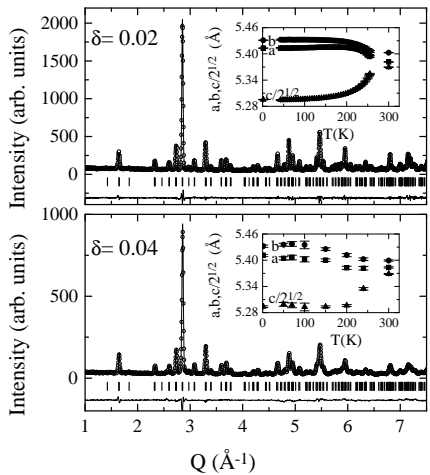,width=0.7\columnwidth}}
\caption{Refined ND patterns at RT of both samples (D2B for
$Nd_{0.5}Ca_{0.5}MnO_{3.02(1)}$ and 3T2 for
$Nd_{0.5}Ca_{0.5}MnO_{3.04(1)}$). Inset: evolution with temperature of
the lattice parameter for $Nd_{0.5}Ca_{0.5}MnO_{3.02(1)}$ (using D1B)
and $Nd_{0.5}Ca_{0.5}MnO_{3.04(1)}$ (using G4.2).}
\label{Fig1}
\end{figure}

Figure \ref{Fig1} shows the high-resolution ND patterns, collected at
RT, refined using an average $P\,bnm$ structure. The refined lattice
parameters, $Mn$-$O$ bond lengths and $Mn$-$O$-$Mn$ bond angles are
listed in Tab.~\ref{Tab1} for comparison. There is a good agreement
between the structural parameters obtained for $Nd_{0.5} Ca_{0.5}
MnO_{3.02(1)}$ with those for $Nd_{0.5} Ca_{0.5} MnO_{3.04(1)}$ and
with those previously reported for $Nd_{0.5} Ca_{0.5}
MnO_3$.\cite{Woodward98} The lattice parameters show a negligible
compression of the $c$-axis. The $MnO_6$ octahedra are very regular
and without any appreciable apical compression. In order to check the
mean oxidation state of the $Mn$ ions using ND data, we refined the
occupation factors of $Nd$, $Ca$ and $Mn$ ions assuming that no
interstitial oxygens are present (that is, fixing the oxygen
stoichiometry to 3). It is known that any excess of oxygen in
perovskites always corresponds to the presence of cation vacancies
since there is no space left in the perovskite structure to allocate
interstitial oxygens. We observed no improvement of the refinement
factors and no significant differences between the nominal composition
and the refined values: $Nd_{0.502(8)}Ca_{0.492(8)}Mn_{1.001(9)}O_{3}$
for the air sample $Nd_{0.496(8)}Ca_{0.492(8)}Mn_{0.992(8)}O_{3}$ for
the $O_2$ sample.

\begin{figure}[t]
\centerline{%
\epsfig{file=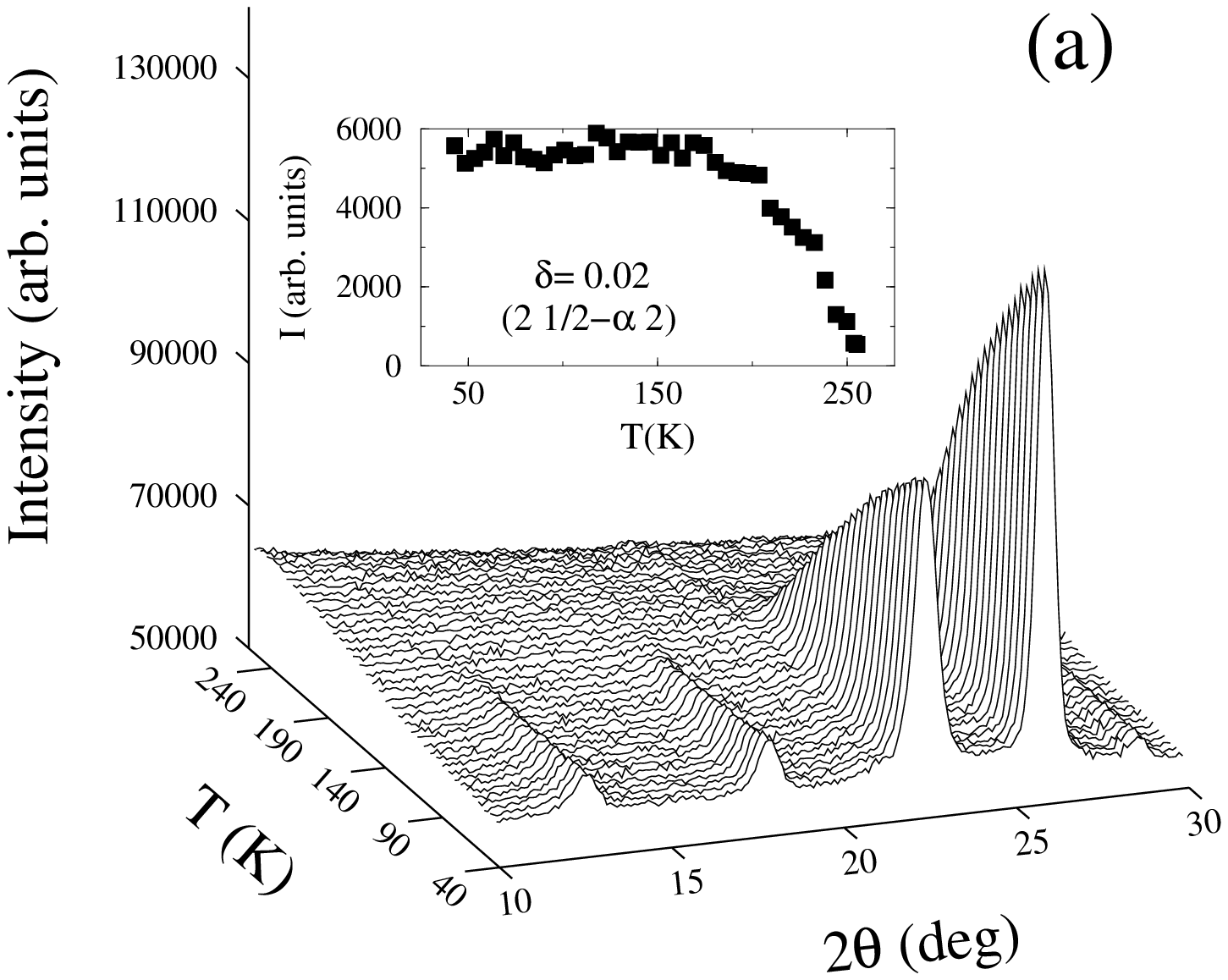,width=0.65\columnwidth}%
}%
\centerline{\epsfig{file=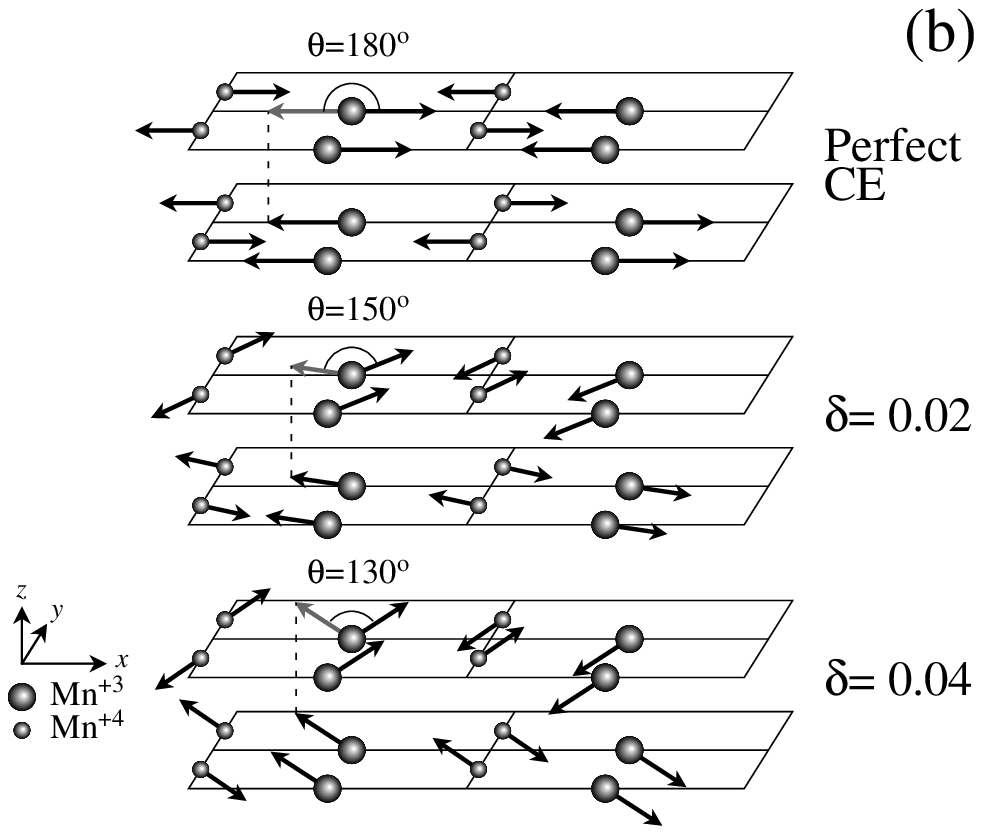,width=0.65\columnwidth}}
\caption{(a) Low angle region of the collected ND patterns showing the onset 
of the AFM order. Inset: Integrated
intensity of the $(2\; 1/2\!-\!\alpha\; 2)$ supperlattice peak showing the
CO/OO transition. Both correspond to $Nd_{0.5}Ca_{0.5}MnO_{3.02(1)}$
data collected at D1B diffractometer.  (b) Scheme of the magnetic
structures obtained for $Nd_{0.5}Ca_{0.5}MnO_{3.02(1)}$ (middle) and
$Nd_{0.5}Ca_{0.5}MnO_{3.04(1)}$ (bottom) compared to that of the
perfect CE-type order (top).  The arrows in grey of the upper planes
correspond to the translation of the magnetic moment of the $Mn$
placed just bellow, in order to visualize the angle between the two
magnetic moments. }
\label{Fig2}
\end{figure}

ND patterns of both samples present superlattice peaks below
$T_{CO}\approx 250K$, attributed to the development of the CO/OO
state. A small incommensurability of the charge order can be
appreciated through the position of the  $(2\;1/2\!-\!\alpha\;2)$ peak
for the $Nd_{0.5}Ca_{0.5}MnO_{3.04(1)}$ sample [$\alpha=0.03(1)$] but
not, within  the resolution of our ND data, for the
$Nd_{0.5}Ca_{0.5}MnO_{3.02(1)}$ sample ($\alpha \approx 0$). The inset
of Fig.~\ref{Fig2}(a) shows the growth of the $(2\;1/2\!-\!\alpha\;2)$
peak found for $Nd_{0.5}Ca_{0.5}MnO_{3.02(1)}$. Its integrated
intensity relative to that of the $(0\;2\;2),\;(2\;0\;2)$ positions
is, at low temperature,
$I_{(2\;1/2-\alpha\;2)}/I_{(0\;2\;2),\;(2\;0\;2)}= 7.4\,10^{-3}$ for
$Nd_{0.5}Ca_{0.5}MnO_{3.02(1)}$ and $7.1\,10^{-3}$ for
$Nd_{0.5}Ca_{0.5}MnO_{3.04(1)}$, indicating a very similar degree of
long range ordering in both samples. The effect on the lattice
parameters of the CO/OO is shown in the insets of Fig.~\ref{Fig1}. The
strong compression of the $c$ lattice parameter is due to the
localization of the $e_g$ electrons in $d_{3x^2-r^2}$ or
$d_{3y^2-r^2}$ orbitals contained in the $(0\;0\;1)$ planes. The low
temperature ND patterns have been refined using the $P\,bnm$
description of the single cell. The obtained lattice parameters,
$Mn$-$O$ bond distances and $Mn$-$O$-$Mn$ bond angles are listed in
Tab.~\ref{Tab1}.  The aforementioned in-plane localization of the
$e_g$ electrons causes an apical compression of the $MnO_6$ octahedra
that can be quantified through the parameter $\epsilon_d=\left
|1-\frac{d_{Mn-O1}}{\langle d_{Mn-O2} \rangle} \right| \times
10^4$. $\epsilon_d$ is, at low temperature, about four times larger
than at RT for both samples. $La_{1/2}Ca_{1/2}MnO_3$
($\epsilon_d=205$)\cite{Radaelli97} and $Nd_{1/2}Sr_{1/2}MnO_3$
($\epsilon_d= 215$)\cite{Kajimoto99}, both presenting CO and a CE-type
magnetic structure, present an apical compression comparable to
$Nd_{0.5}Ca_{0.5}MnO_{3.02(1)}$ ($\epsilon_d=217$). A little lower
value is found for $Nd_{0.5}Ca_{0.5}MnO_{3.04(1)}$ ($\epsilon_d=192$),
reflecting the partial lack of $e_g$ electrons.

\begin{figure}
\centerline{\epsfig{file=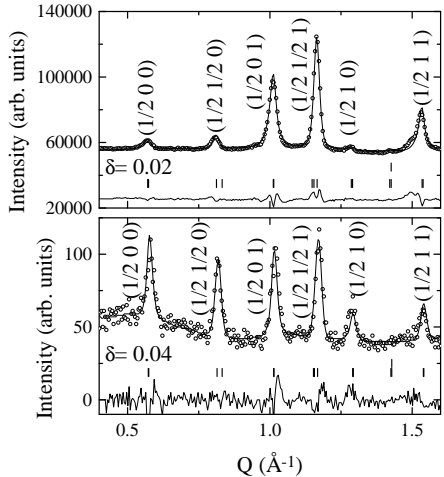,width=0.7\columnwidth}}
\caption{Detail of the low temperature refined patterns showing the
low angle region with the most intense magnetic peaks. $\lambda=
2.52{\rm \AA}$ (D1B) has been used for $Nd_{0.5}Ca_{0.5}MnO_{3.02(1)}$
and $\lambda= 2.426{\rm \AA}$ (G4.2) for
$Nd_{0.5}Ca_{0.5}MnO_{3.04(1)}$.}
\label{Fig3}
\end{figure}

Figure \ref{Fig2}(a) shows the temperature evolution of the low angle
region of the ND patterns of $Nd_{0.5}Ca_{0.5}MnO_{3.02(1)}$. The
apparition of magnetic reflections is observed to occur at $T_{N1}=
165K$ [$Nd_{0.5}Ca_{0.5}MnO_{3.02(1)}$] and $T_{N2}= 150K$
[$Nd_{0.5}Ca_{0.5}MnO_{3.04(1)}$]. The low angle regions of the
refined ND patterns at low temperature are plotted in Fig.~\ref{Fig3}
for both samples. The magnetic peaks have been indexed on the basis of
the $P\,bnm$ setting. In both samples, there are two families of
magnetic peaks displaying $(h\; k\; l)$ Miller indices with $h$ half
integer and k both integer or half integer. The first family of peaks
has $l=$odd [$(1/2\; 0\; 1)$, $(1/2\; 1/2\; 1)$,\dots] and is
associated to the CE-type AFM structure.\cite{Wollan55,Radaelli97} The
second family of peaks has $l=$even [$(1/2\; 0\; 0)$, $(1/2\; 1/2\;
0)$,\dots] and is usually associated to the {\em pseudo CE-type} AFM
structure.\cite{Nota2} These two families of peaks are also present in
$Nd_{0.45}Ca_{0.55}MnO_3$.\cite{Li99} We have obtained good fits to
the ND patterns assuming that the components of the magnetic moment
contained in the $(0\; 0\; 1)$ planes ($m_x$ and $m_y$) present a
CE-type order structure, while the component of the magnetic moment in
the $[0\; 0\; 1]$ direction ($m_z$) presents a {\em pseudo CE-type}
AFM order. The obtained magnetic structures are described in
Tab.~\ref{Tab2} and schematically plotted in Fig.~\ref{Fig2}(b). In
both cases the ordered moments are well below the expected values for
perfectly ordered moments. The effect of the considered out-of-plane
component of the magnetic moment can be interpreted as a deviation
from the perfect value ($180^o$) of the angles formed by the magnetic
moments in neighboring $(0\;0\;1)$ planes [e.g.~between the magnetic
moments of the $Mn^{+3}$ ions in positions $\left (\frac 14,\frac
12,0\right)$ and $\left (\frac 14,\frac 12,\frac 12\right)$ of
Tab.~\ref{Tab2} as is schematically depicted in Fig.~\ref{Fig2}(b)]
that is about $30^o$ in $Nd_{0.5}Ca_{0.5}MnO_{3.02(1)}$ and $50^o$ in
$Nd_{0.5}Ca_{0.5}MnO_{3.04(1)}$.

\begin{figure}
\centerline{\epsfig{file=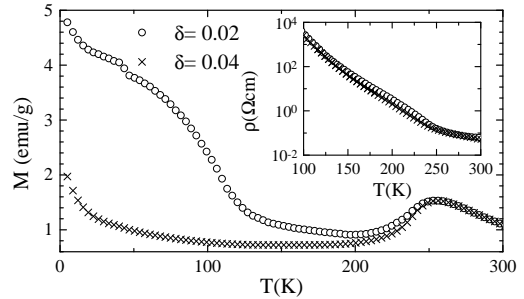,width=0.8\columnwidth}}
\caption{Temperature evolution of the magnetization M(T) (applied
field $0.5T$) and $\rho(T)$ (inset) for both
$Nd_{0.5}Ca_{0.5}MnO_{3.02(1)}$ and $Nd_{0.5}Ca_{0.5}MnO_{3.04(1)}$.}
\label{Fig4}
\end{figure}

The magnetization [M(T)] and resistivity [$\rho(T)$] curves are shown
in Fig.~\ref{Fig4} and its inset respectively. The CO transition can
be well appreciated as a sudden increase in the $\rho(T)$ curves,
caused by the localization of the charges, and as a local maximum in
the $M(T)$ curves. This local maximum is due to the transition from
the FM $Mn$-$Mn$ correlations above $T_{CO}$ (double-exchange) to the
AFM $Mn$-$Mn$ correlations below $T_{CO}$ ($Mn$-$O$-$Mn$
superexchange). There is scarcely any variance in $\rho(T)$ from one
sample to the other. In this sense, the magnetization curves clearly
show two regimes: above $T_{CO}$ both curves are almost identical, but
below this point the curves systematically start to come away.


The results presented in the previous paragraphs show that the two
samples are indeed very similar: the different oxydation state of the
$Mn$ does not significantly change the structure, the transport and
the magnetic properties of the compound above $T_{CO}$. Consequently,
the differences observed at lower temperatures cannot be attributed to
intrinsic differences in the structure due to the different
concentration of cation vacancies.

In contrast with the behavior above $T_{CO}$, substantial
discrepancies in the $M(T)$ curves start at this temperature, just
when the $Mn$-$O$-$Mn$ superexchange interactions are dominated by the
real space location of the $e_g$ electrons. This indicates that the
difference in the magnetic behavior of the samples is driven by the
defects introduced by the lack of $e_g$ electrons in the OO state.
The different density of carriers causes differences in the real space
distribution of $Mn$-$O$-$Mn$ superexchange interactions giving rise
to a dissimilar $M(T)$ evolution. The presence of defects in the OO
structure gives rise to frustrated magnetic coupling and, hence,
differences between the magnetic structure and the perfect CE-type. To
be emphasized is that such great changes in the obtained magnetic
order and in the $M(T)$ are hard to explain in the scenario of a
random location of the holes among $Mn$ positions. The last will
simply lead to a CE structure with disorder. Consequently, a certain
spatial grouping of the defects may enhance their effects and also
cause the small incommensurability of the CO.

Of special interest is to determine if the two families of magnetic
peaks are sustained by the same structural lattice or they  correspond
to two slightly different lattices. In $Nd_{0.5}Sr_{0.5}MnO_3$, the
coexistence of an A-type and a CE-type AFM phases below the  CO
transition, supported by different crystallographic cells, has been
attributed to inhomogeneities in the cation
distribution.\cite{Woodward99} In the present case (for both samples)
the two families of magnetic peaks can be very well reproduced using
the same lattice parameters and appear at the same temperature [see
Fig.~\ref{Fig2}(a)]. Within our resolution, a  single set of cell
parameters allows a proper indexation of all  the magnetic intensities
(see Fig.~\ref{Fig3}).

In summary, off-stoichiometric samples of $Nd_{0.5}Ca_{0.5}MnO_{3.02(1)}$
[$\%Mn^{+4}=54(2)$] and $Nd_{0.5}Ca_{0.5}MnO_{3.04(1)}$
[$\%Mn^{+4}=58(2)$] samples have been investigated in comparison
with the stoichiometric $Nd_{0.5}Ca_{0.5}MnO_3$. The same structural
features and macroscopic behavior are observed above $T_{CO}$. A
detailed ND study reveals pronounced influence of the different, but
very similar, oxidation state of $Mn$ upon the magnetic long range
ordering. The intensities of the magnetic
$(h/2\;k/2\;0)$ peaks (absent in the CE-type order) increase
significantly with the value of $\delta$. As a result the angle formed
by the spins of successive $MnO_2$ $(0\;0\;1)$ planes change from
$\theta=180^o$ for $\delta=0$ to $\theta=130^o$ for
$\delta=0.04$. These remarkable changes in the magnetic structure are
unusually important compared to the effects of a small
off-stoichiometry in most of the transition metal oxides.  They
indicate, in connection with the incommensurability of the CO detected
for the $\delta=0.04(1)$ sample, that the extra holes are not randomly
placed in the $Mn$ positions but they form spatial sub-structures.


This work has been done with financial support from the CICyT
(MAT97-0669) MEC (PB97-1175) and Generalitat de Catalunya
(GRQ95-8029). A.L. acknowledges financial support form the Oxide Spin
Electronics Network (EU TMR) program. The ILL and LLB are acknowledged
for making available the beam time.

\newpage

\begin{table}[p]
\caption{Refined structural parameters of
$Nd_{0.5}Ca_{0.5}MnO_{3+\delta}$ at room and low
temperature. Parameter $\epsilon_d$ is defined in the text. O1 stands
for apical and O2 for basal oxygens. }
\label{Tab1}
\begin{tabular}{l||c|c||c|c}
&\multicolumn{2}{c||}{$\delta= 0.02(1)$ (D2B)}
&\multicolumn{2}{c}{$\delta= 0.04(1)$ (3T2)}\\ \hline &4K&RT&1.5K&RT
\\ \hline $a$ (\AA)&5.4133(6)&5.3821(4)&5.4055(7)&5.3771(6)\\ $b$
(\AA)&5.4326(6)&5.4038(4)&5.4259(7)&5.4010(6)\\ $c$
(\AA)&7.4927(7)&7.5923(4)&7.4807(8)&7.5874(8)\\ $V$
($\AA^3$)&220.35&220.81&219.40&220.35\\
$\sqrt{2}c/(a+b)$&0.977&0.995&0.977&0.996\\ $\langle
d_{Mn-O1}\rangle$&1.9106(9)&1.9354(6)&1.9142(9)&1.933(1)\\ $\langle
d_{Mn-O2}\rangle$&1.953(3)&1.946(2)&1.952(4)&1.945(4)\\ $\langle
d_{Mn-O}\rangle$&1.939(2)&1.942(2)&1.939(3)&1.941(3)\\
$\epsilon_d$&217(14)&52(9)&192(18)&59(15)\\ 
$\theta_{Mn-O1-Mn} (deg.)$&157.3(1)&157.4(1)&155.5(1)&157.8(1)\\ 
$\theta_{Mn-O2-Mn} (deg.)$&158.1(1)&157.0(1)&157.9(2)&156.9(2)\\ $\langle
\theta_{Mn-O-Mn}\rangle
(deg.)$&157.8(1)&157.2(1)&157.8(2)&157.2(2)\\
$\chi^2$&3.6&2.3&3.8&2.8\\ $R_N(\%)$&8.0&4.8&8.5&8.3

\end{tabular}
\end{table}

\mediumtext
\begin{table}
\caption{Magnetic structures obtained at low temperature for
$Nd_{0.5}Ca_{0.5}MnO_{3.02(1)}$ and $Nd_{0.5}Ca_{0.5}MnO_{3.04(1)}$.
The atomic positions are referred to the magnetic lattice ($2a\times
2b\times c$). }
\label{Tab2}
\begin{tabular}{c|c||c|c|c|c||c|c|c|c}
& Position & \multicolumn{4}{c||}{\begin{tabular}{c}
 $Nd_{0.5}Ca_{0.5}MnO_{3.02(1)}$ (D1B) \\
 ($R_{mag}=8\%$)\end{tabular}} & \multicolumn{4}{c}{\begin{tabular}{c}
 $Nd_{0.5}Ca_{0.5}MnO_{3.04(1)}$ (G4.2) \\
 ($R_{mag}=7\%$)\end{tabular}}\\ \hline & & $m_x(\mu_B)$ &
 $m_y(\mu_B)$ &$m_z(\mu_B)$ & $m_T(\mu_B)$ & $m_x(\mu_B)$ &
 $m_y(\mu_B)$ &$m_z(\mu_B)$ & $m_T(\mu_B)$ \\ \hline $Mn^{+3}$ &
\begin{tabular}{c} $\left (\frac 14, 0, 0\right)$;
$\left (\frac 14, \frac 12, 0\right)$\\ 
$\left (\frac 14, 0, \frac 12\right)$; 
$\left (\frac 14, \frac 12, \frac 12\right)$ \\
$\left (\frac 34, 0, 0\right )$; 
$\left (\frac 34, \frac 12, 0\right )$\\ 
$\left (\frac 34, 0, \frac 12\right)$; 
$\left (\frac 34, \frac 12, \frac 12\right)$
\end{tabular} &
\begin{tabular}{r} 1.9(1)\\-1.9(1)\\-1.9(1)\\1.9(1)\end{tabular} &
\begin{tabular}{r} 1.9(1)\\-1.9(1)\\-1.9(1)\\1.9(1)\end{tabular} &
\begin{tabular}{r} 0.7(2)\\0.7(2)\\-0.7(2)\\-0.7(2)\end{tabular} & 2.8(2)&
\begin{tabular}{r} 2.5(2)\\-2.5(2)\\-2.5(2)\\2.5(2)\end{tabular} &
\begin{tabular}{r} 0\\0\\0\\0\end{tabular} &
\begin{tabular}{r} 1.5(1)\\1.5(1)\\-1.5(1)\\-1.5(1)\end{tabular} & 3.0(2)\\
\hline $Mn^{+4}$ &
\begin{tabular}{c} 
$\left (0, \frac 14, 0\right)$;
$\left (\frac 12, \frac 34, 0\right)$\\
$\left (0, \frac 14, \frac 12\right)$; 
$\left (\frac 12, \frac 34, \frac 12\right)$\\
$\left (0, \frac 34, 0\right)$; 
$\left (\frac 12, \frac 14, 0\right)$\\ 
$\left (0, \frac 34, \frac 12\right)$; 
$\left (\frac 12, \frac 14, \frac 12\right)$\\
\end{tabular} &
\begin{tabular}{r} 1.7(1)\\-1.7(1)\\-1.7(1)\\1.7(1)\end{tabular} &
\begin{tabular}{r} 1.7(1)\\-1.7(1)\\-1.7(1)\\1.7(1)\end{tabular} &
\begin{tabular}{r} 0.7(2)\\0.7(2)\\-0.7(2)\\-0.7(2)\end{tabular} & 2.5(2)&
\begin{tabular}{r} 1.9(2)\\-1.9(2)\\-1.9(2)\\1.9(2)\end{tabular} &
\begin{tabular}{r} 0\\0\\0\\0\end{tabular} &
\begin{tabular}{r} 1.3(1)\\1.3(1)\\-1.3(1)\\-1.3(1)\end{tabular} & 2.5(1)

\end{tabular}
\end{table}
\narrowtext
\clearpage


\begin{references}


\bibitem{Radaelli97}P.G. Radaelli, D.E. Cox, M. Marezio, and
S.-W. Cheong, Phys.~Rev.~B {\bf 55}, 3015 (1997).

\bibitem{Wollan55}E.O. Wollan and W.C. Koehler, Phys. Rev. {\bf 100},
545 (1955). J.B. Goodenough, Phys. Rev. {\bf 100}, 564 (1955).

\bibitem{Roy98}M. Roy, J.F. Mitchell, A.P. Ramirez and P. Schiffer,
Phys. Rev. B {\bf 58}, 5185 (1998).

\bibitem{Tomioka96}Y. Tomioka, A. Asamitsu, H. Kuwahara, Y. Moritomo
and Y. Tokura, Phys. Rev. B {\bf 53}, R1689 (1996).

\bibitem{Barnabe98}A. Barnab\'e, M. Hervieu, C. Martin, A. Maignan,
and B.  Raveau, J. of Appl. Phys. {\bf 84}, 5506 (1998).

\bibitem{Chen99}C.H. Chen, S. Mori, and S-W. Cheong,
Phys. Rev. Lett. {\bf 83}, 4792 (1999).

\bibitem{Richard99}O. Richard, W. Schuddinck, G. Van Tendeloo,
F. Millange, M. Hervieu, V. Caignaert, and B. Raveau, Acta
Cryst. A{\bf 55}, 704 (1999).

\bibitem{Chen96}C.H. Chen and S-W. Cheong,
Phys. Rev. Lett. {\bf 76}, 4042 (1996).

\bibitem{JRC93}J. Rodr\'{\i}guez-Carvajal, Physica B {\bf 192} (1993)
55.

\bibitem{Woodward98}P.M. Woodward, T. Vogt, D.E. Cox, A. Arulraj,
C.N.R. Rao, P. Karen, and A.K. Cheetham, Chem. Mater. {\bf 10}, 3652
(1998).

\bibitem{Kajimoto99}R.Kajimoto, H. Kawano, H. Kurwahara, Y. Tokura,
K. Ohoyama and M. Ohashi, Phys. Rev. B {\bf 60}, 9506 (1999).

\bibitem{Nota2}AFM $(0\; 0\; 1)$ planes coupled ferromagnetically [see
Z. Jir\'ak, S. Krupi\v{c}ka, Z. \v{S}im\v{s}a, M.  Dlouh\'a, and
S. Vratislav, J.~Magn.~Magn.~Mater.~{\bf 53}, 153 (1985)].

\bibitem{Li99}W.-H. Li, S.Y. Wu, K.C. Lee, J.W. Lynn, R.S. Liu,
J.B. Wu, and C.Y. Huang, J. App. Phys. {\bf 85}, 5588 (1999).

\bibitem{Woodward99}P.M. Woodward, D.E. Cox, T. Vogt, C.N.R. Rao and
A.K. Cheetham, Chem. Mater. {\bf 11}, 3528 (1999).


\end{references}
\end{document}